\documentclass[12pt]{article}
\usepackage{amssymb,amsfonts}
\newcommand{\Lm}{\Lambda_{-}}
\newcommand{\Lp}{\Lambda_{+}}

\newcommand{\ddF}{\delta F}
\newcommand{\ddG}{\delta G}
\newcommand{\ddg}{\delta {\bf g}}

\newcommand{\ddp}{\delta {\bf p}}
\begin{document}

\centerline{\bf Darboux Transformation and Exact Solutions}
\centerline{\bf in the model of Cylindrically Symmetrical Chiral Field}
\vskip0.4cm
\hfil{\bf E.Sh.Gutshabash}\hfil
\vskip0.4cm
\hfil{St.-Petersburg State University, Russia}

\hfil{e-mail: gutshab@EG2097.spb.edu}
\vskip0.4cm

\hskip 2cm \parbox{13cm} {\small The application of the Darboux Transformation method to the integrable
 model of Cylindrically Symmetrical Chiral field has been considered.
 The associated linear system of matrix equations has been proposed
 and the properties of symmetrie for its solutions has been obtained.
 The necessary form of Darboux Transformation has been found and
 formal one- and $N$-soliton solutions have been constructed. With the use of
 Pohlmayer's Transformation the equation $\sin$-Gordon type have been
 given and the hypothesis about its integrability has been deduced.}
\vskip0.8cm

{\bf 1.} It is well known, that all methods of solution of nonlinear integrable
equations of modern theoretical and mathematical physics can be
divided into two large groups: the direct, when the Lax
representation of initial equation is not required, and indirect one, when we have
the evident representation of our nonlinear equation as the overdetermined
system two (matrix) linear equations on any auxiliary function.

To the first group one can assign the ansatz-method,
the Hirota's method and etc. The second group includes such methods  as the Backlund
transformation, the different versions of the dressing procedures, the
Inverse Scattering Transform. Among this group
the method of Darboux transformation (DT) has a significant place. The numerous of
examples of its advantegeoures use are given in [1].
However the further progress of the theory of nonlinear
integrable equations requires that the new and more
complex versions of DT are involved .

The aim of this paper is to demonstrate one of this versions by the
example of equation cylindrically symmetrical chiral field ($\sigma$-
model), considered originally in [2]. In this case
the method of DT using appear to be rather nontrivial, it calls for a
preliminary detailed study of the algebraic properties of the Lax representation.

{\bf 2.} Equation of the model of the cylindrically symmetrical chiral
field has the form:

$$\partial_{\eta}(\alpha g_{\xi}g^{-1})+\partial_{\xi}(\alpha
g_{\eta}g^{-1})=0 ,\eqno(1)
$$
where $\xi=(t-r)/2,\; \eta=(t+r)/2$ are the cone variables, $g$ is the element
of some group Lee ${\cal G}$ (or symmetrical spaces), $\alpha=r$.

This equation can be rewritten in two equivalent forms:

$$
(g_tg^{-1})_t=(g_rg^{-1})_r+\frac {1}{r}g_rg^{-1} , \eqno(1')
$$
$$
(rg_tg^{-1})_t=(rg_rg^{-1})_r . \eqno(1'')
$$

It should be noted, that appearance of equation (1) is coincides with one of the
following equations: one
of the versions of the Ernst's equations in the gravitation' theory [3]
(for ${\cal G}=
SL(2)/SO(2))$, the model of selfdual Yang-Mills's field (model
of Zakharov-Manakov) [4] (for ${\cal G}=SU[N]$), the model of stationary
two-dimensional ferromagnet of Heisenberg possessing variable nominal
magnetization [5, 6]; the appearance is encountered in the series of another problems.
The
significance of the equation (1) is that it is the preimidge more realistic
four-dimensional gauge theories.

In [2] this equation was studied to perfect the procedure dressing
solutions. Also in paper [7] it has been examined employing the
method of the harmonical maps.

It should be notice, that the model (1) becomes (2+1)-dimensional
when it is supposed that $r=\sqrt{x^2+y^2}$,
where $x,\:y$ are the Cartesian coordinates on the plane.

Here we will restrict ourselves to the case of $g \in GL(2,{\mathbb C})$,
with $g$ can be parameterized in the form:
$g=\sum_{i=1}^
3 g_i\sigma_i,
\: {\bf g}=(g_1, g_2, g_3)$ is the unit vector with the real components,
$\sigma_i$ are the standard Pauli matrices. Then $g$ posses the properties:
$$
g^2=I,\qquad \det g=-1,\qquad g=g^{+}, \qquad Tr\; g=0, \eqno(2)
$$
where $I$ is unit $2\times 2$ matrices, and symbol $g^{+}$ means the
hermitian conjugate.

At first we will dwell on the linear version of equation (1).
In this case it reduces to:

$$
g_{tt}=g_{rr}+\frac {1}{r} g_{r} , \eqno(3)
$$
and can be easy solved. We have

$$
g=\left(\begin{array}{cc}
g_3&g_{-}\\
g_{+}&-g_{3} \end{array}\right) ,\eqno(4)
$$
where

$$
g_i(r,t)= \int dk (C_i^1(k)e^{kt}+C_i^2(k)e^{-kt})
(a_i(k) J_0(kr)+b_i(k) N_0(kr)) , \eqno(5)
$$
and $g_{+}=g_1+ig_2,\; g_{-}={\bar g_{+} }$. Here $J_0,\:N_0$
are the function of Bessel and Neeman, functions $C_i^{1},\: C_i^{2},\: a_i,\:
b_i$ should be determined from the initial and boundary conditions, and also
the condition $\det g=-1$.

Returning to the nonlinear equation (1), one can attract the ansatz of
the following specials kind:

$$
g=\left(\begin{array}{cc}
\cos \chi&e^{-i\Phi}\sin \chi\\
e^{i\Phi}\sin \chi&-\cos \chi \end{array}\right) ,\eqno(6)
$$
where $\Phi=\Phi (r,t),\: \chi=\chi (r,t)$ are real functions, which need to
determination. Substituting (6) into (1), after simple calculations we
obtain the system of nonlinear differential equations:

$$
(\Phi_{tt}-\Phi_{rr}-\frac {1}{r}\Phi_r)\sin \chi+2(\Phi_t\chi_t-\Phi_r\chi_r)
\cos \chi=0 ,
$$
$$
\eqno(7)
$$
$$
2(\chi_{tt}-\chi_{rr}-\frac {1}{r}\chi_r)=(\Phi_t^2-\Phi_r^2)\sin 2\chi .
$$

There is interest some special cases of this system. Let
$\chi=0$, then, evidently, $g=\sigma_3$. At $\chi=\pi/2$ on the
function $\Phi$ we obtain the scalar equation, which coincides on
his kind with (3). The same equation arises at $\Phi=0$ on
function $\chi$. Now setting $\Phi=kt$, where $k$ is the free
parameter, from the first equation (7) it follows that $\chi=\chi
(r)$, and second equation gives:

$$
\chi_{rr}+\frac {1}{r}\chi_r=-\frac {k^2}{2}\sin 2\chi . \eqno(8)
$$
This equation at the properly imagine
$k$ after the automodel substitution reduces to the equation Painleve-III.

{\bf 3.} To apply DT method to (1) we write the associated linear
system of equation as

$$
\Psi_{\xi}=U\Psi\Lm \;\;,\;\;
\Psi_{\eta}=V\Psi\Lp ,\eqno(9)
$$
where $\Psi(\xi,\eta,\lambda)\in Mat(2,{\mathbb C}),\; U=g_{\xi}g^{-1},\;
V=g_{\eta}g^{-1}$ are left currents (elements algebras Lee  ${\rm g}$
), and diagonal matrices
$\Lm,\: \Lp$, are determined by the expressions:

$$
\Lm=\left(\begin{array}{cc}
-\frac {r}{\gamma-r}&0\\
0&\frac {{\overline {\gamma}}}{\overline {\gamma}-r} \end{array}\right) ,
\qquad
\Lp=\left(\begin{array}{cc}
\frac{r}{\gamma+r}&0\\
0&\frac {{\overline {\gamma}}}{\overline {\gamma}+r}\end{array}\right) .
\eqno(10)
$$
The function $\gamma(\xi,\eta,\lambda) \equiv \gamma(r,t,\lambda)$ which was
introduced in (3), contains so-called "hidden" complex parameter:

$$
\gamma=\xi+\eta-\lambda+\sqrt{(\lambda-2\xi)(\lambda-2\eta)}=t-\lambda+
\sqrt{(t-\lambda)^2-r^2} ,\eqno(11)
$$
where $\lambda \in {\mathbb C}$ is the parameter that does not depend on the
coordinate and time.

The condition of the compatibility of the system (2)
: $\Psi_{\xi\eta}=\Psi_{\eta\xi}$ is equivalent to both of following two relations:

$$
U_{\eta}-V_{\xi}+[U,V]=0 ,
$$
which asserts identically and (1). So, equation (1) is quite integrable.

We will study some properties of system (2), including the symmetries.

Matrices $\Lp,\: \Lm$ satisfy the series of useful identities. In order to
demonstrate this fact, we introduce also the matrices
$\sigma_0$ and $M$:

$$
\sigma_0=\left(\begin{array}{cc}
\frac {\gamma}{(\gamma+r)(\gamma-r)}&0\\0& -\frac {\overline {\gamma}}
{(\overline {\gamma}-r)(\bar {\gamma}+r)})\end{array}\right)   ,\eqno(12)
$$
\vskip0.3cm
$$
M=\sigma_0\Lp^{-1}\Lm^{-1}= \left(\begin{array}{cc}
\frac {r^2}{\gamma}&0\\
0&{{\overline {\gamma}}} \end{array}\right)  .\eqno(13)
$$
By the direct calculation one can check, that the following identities are true:

$$ \Lp=\Lp\Lm-r\sigma_0,\qquad  \Lambda_{-\eta}=\sigma_0 ,
\eqno(14) $$ $$ \Lm=\Lp\Lm+r\sigma_0,\qquad
\Lambda_{+\xi}=\sigma_0 ,  \eqno(15) $$ $$ M\Lp=M-r\Lp,\qquad
M_{\eta}=2\Lp,                \eqno(16) $$ $$ M\Lm=M+r\Lm,\qquad
M_{\xi}=2\Lm  ,                 \eqno(17) $$ $$ \Lp=I+rM^{-1}
,\qquad  \Lm=I+rM^{-1} ,\eqno(18) $$ $$ \Lambda_{-\xi}=-\frac
{2}{r}\Lm^3+\frac {3}{r}\Lm^2-\frac {1}{r}\Lm ,\eqno(19) $$ $$
\Lambda_{+\eta}=\frac {2}{r}\Lp^3-\frac {3}{r}\Lp^2+\frac
{1}{r}\Lp.\eqno(20) $$ Series of relations (14)-(20) it is
naturally to call the kinematical connections. Using them it is
not difficult to obtain some more useful two connection, written
in terms matrices $M$:

$$
M_{\xi\xi}=-\frac {1}{2r}M_{\xi}^3+\frac {3}{2r}M_{\xi}^2-\frac{1}{r}M_{\xi}  ,
$$
$$
\eqno(21)
$$
$$
M_{\eta\eta}=\frac {1}{2r}M_{\eta}^3-\frac {3}{2r}M_{\eta}^2+
\frac{1}{r}M_{\eta}  .
$$
Moreover, for matrices  $M$ we have:

$$
M_{\xi\eta}=-2(I-rM^{-1})^{-1}(I+rM^{-1})M^{-1}  , \eqno(22)
$$
that is it satisfy to nonlinear wave equation and

$$
\left.\left [M, M_{\xi} \right.\right ]=\left.\left [M, M_{\eta}
\right.\right ]=\left.\left [M_{\xi},
M_{\eta} \right.\right ]=\left.\left [M_{\xi\xi}, M_{\eta\eta}
\right.\right]=\cdots =0 .\eqno(23)
$$

Now we will study the properties of symmetry of the solution of the equation (1).
From the determination matrices
$g$
it follows, that the current must meet the condition:

$$
U=\sigma_2\overline U\sigma_2  .              \eqno(24)
$$
Let ${\hat \tau}$ - the operation of the transposition of the sheets of
Riemann surface
$\Gamma$ for function $\gamma: \gamma(\hat \tau(\lambda))=r^2/\gamma(\lambda)$.
Then

$$
\Lambda_{\pm}(\hat \tau(\lambda))=\sigma_2\overline
{\Lambda}_{\pm}(\lambda)\sigma_2
.\eqno(25)
$$
From the expressions (9), (24)-(25) it follows

$$
\Psi(\lambda)=\sigma_2\overline {\Psi}(\hat \tau(\lambda))\sigma_2 .\eqno(26)
$$
Since $g=g^{-1}$, and taking into account (9), we obtain:

$$ \sigma_2g\Psi(\lambda)\sigma_2=\overline
{\Psi}(\lambda)\sigma_3 . \eqno(27)
$$
Relations (25)-(26), which
are the consequence of the properties of function $ \gamma(\xi,
\eta)$, are the whole set of the involutions of the problem
(9), so that all solutions determined below must be satisfy it.

{\bf 4.} Let's consider the construction the exact solutions of the model (1).
In view of,  both equation of the Lax pair should be covariant relatively DT and
taking
into account the second equality (16)-(17), we rewriting the associated
linear system (9) in the more symmetrical form:

$$
\Psi_{\xi}=\frac {1}{2}U \Psi M_{\xi} ,\qquad
\Psi_{\eta}=\frac {1}{2}V \Psi M_{\eta} .\eqno(28)
$$

Check the covariance (28) relatively the matrix DT:

$$
\tilde \Psi=\Psi-L_1 \Psi M ,\eqno(29)
$$
where $L_1=\Psi_1M_1^{-1}\Psi_1^{-1}$,   $\Psi_1$ is some fixed
solution of equation (9). So, we require that (9) holds its form:

$$
\tilde {\Psi}_{\xi}=\frac {1}{2}\tilde U \tilde{\Psi} M_{\xi},\;\;\;\;
\tilde {\Psi}_{\eta}=\frac {1}{2}\tilde V \tilde{\Psi} M_{\eta} ,\eqno(30)
$$
where $\tilde U=
\tilde {g}_{\xi}\tilde {g}^{-1},\; \tilde {V}=\tilde {g}_{\eta}
\tilde {g}^{-1}$ are the  "dressing" currents.

Substituting the ansatz (29) in the first equation of the system (30),
we find:

$$
\tilde U=U+rL_{1\xi}-2L_1 ,   \eqno(31)
$$
$$
\tilde U=L_1UL_1^{-1}+L_{1\xi}L_1^{-1} ,   \eqno(32)
$$
where $U$ is the initial solution of equation (1). Note also, that the check
of the equivalence this relations gives identity (19).

One can write formulas (31) in more conventional form:

$$
\tilde U=U+r\Psi_1[\Psi_1^{-1}\Psi_{1\xi},M_1^{-1}]\Psi_1^{-1}-
\Psi_1M_1^{-1}M_{1\xi}\Psi_1^{-1}  .\eqno(33)
$$
Analogously, from the second equation (28) we will have:

$$
\tilde V=V-rL_{1\eta}-2L_1 ,  \eqno(34)
$$
$$
\tilde V=L_1VL_1^{-1}+L_{1\eta}L_1^{-1} , \eqno(35)
$$
where $V$ is the initial solution equation (1) and the check of the
equivalence (34) and (35) gives identities (20). Besides,

$$
\tilde V=V-r\Psi_1[\Psi_1^{-1}\Psi_{1\eta},M_1^{-1}]\Psi_1^{-1}-
\Psi_1M_1^{-1}M_{1\eta}\Psi_1^{-1}  .\eqno(36)
$$

But the solutions (33), (36), which are the dressing relations not yet
desired solutions (1). Indeed, from (33) and (36) we have:

$$
Tr \tilde {U} \ne 0 \;\;,\;\;Tr \tilde {V} \ne 0  .\eqno(37)
$$
This inequalities contradict the condition (2). In order to remedy the situation
it should be noted that in equation (1) the currents $U,\: V$ are not
single-valued but are determined with an accuracy of substitution:

$$
U \to U-1/2[\gamma_{\xi}-r(\ln {\overline {\gamma}})_
{\xi}]I, \qquad
V \to V+1/2[\gamma_{\eta}+r(\ln {\overline {\gamma}})_{\eta}]I .\eqno(38)
$$
Therefore, the final expressions for the currents can be written as:

$$
\tilde U=U+r\Psi_1[\Psi_1^{-1}\Psi_{1\xi},M_1^{-1}]\Psi_1^{-1}-
\Psi_1M_1^{-1}M_{1\xi}\Psi_1^{-1}-\frac {1}{2}[-2(\ln r)_{\xi}+
(\ln {\frac {\gamma}{\overline {\gamma}}})_{\xi}]I  ,  \eqno(39)
$$

$$
\tilde V=V-r\Psi_1[\Psi_1^{-1}\Psi_{1\eta},M_1^{-1}]\Psi_1^{-1}-
\Psi_1M_1^{-1}M_{1\eta}\Psi_1^{-1}-\frac {1}{2}[-2(\ln r)_{\eta}+
(\ln {\frac {\gamma}{\overline {\gamma}}})_{\eta}]I . \eqno(40)
$$
Using, for examples (39), one can obtain the formal expression for
the initial matrices $\tilde {g}$:

$$
\tilde {g} \equiv g[1]=T\exp\left.\left(\int^{\xi}
A(\xi,\eta)d\xi\right.\right)g^{(1)} ,   \eqno(41)
$$
where $g^{(1)}$ is initial solution of equation (1),
$A(\xi,\eta)$ is the right-side of relation (39) and symbol
$T$ means
$T$ - ordered exponent. Formulas (41) gives the one soliton solution of the
equation (1).

Now we will construct $N$-soliton solution of the problem. For this it should
be notice, that

$$
\Psi[1]=\Psi-L_1\Psi M ,
$$
$$
\Psi[2]=\Psi[1]-L_2\Psi[1]M=\Psi-(L_1+L_2)\Psi M+L_2L_1\Psi M^2 ,
$$
$$
...................................................... \eqno(42)
$$
$$
\Psi[N]=\Psi+T_1\Psi M+............+T_N\Psi M^N  ,
$$
where $L_i=\Psi_iM_i\Psi_i^{-1},\;M_i=M(\lambda_i) ,\;
\lambda_1,....\lambda_N$ is the set of the fixed complex parameters.
Then the coefficients $T_i,\;
i=1, N$, can be determined from the conditions:

$$ \Psi[N]_{|\Psi=\Psi_i,M=M_i}=0  ,\eqno(43)
$$
which gives us
the system of linear equations:

$$ T_1\Psi_1 M_1+T_2\Psi_1 M_1^2+.....+T_N\Psi_1 M_1^N=-\Psi_1 ,$$
$$ T_1\Psi_2 M_2+T_2\Psi_2 M_2^2+.....+T_N\Psi_2 M_2^N=-\Psi_2 ,$$
$$ ..........................................  \eqno(44) $$ $$
T_1\Psi_N M_N+T_2\Psi_N M_N^2+.....+T_N\Psi_N M_N^N=-\Psi_N .$$
On the another side, accordingly (39)

$$
U[1]=U+rL_{1\xi}-2L_1+Q_1  ,
$$
$$
U[2]=U[1]+rL_{2\xi}-2L_2+Q_2=U+r(L_{1\xi}+L_{2\xi})-2(L_1+L_2)+Q_2 ,
$$
$$
............................................   \eqno(45)
$$
$$
U[N]=U+r\sum_{i=1}^N L_{i\xi}-2\sum_{i=1}^{N} L_i+\sum_{i=1}^N Q_i  ,
$$
where $Q_i=\left.\left [\ln (\gamma(\lambda_i)/r^2
\gamma(\overline {\lambda}_i))\right.\right]_{\xi}$.
Thus, for the $N$th dressed current we will have:

$$
U[N]=U+rT_{1\xi}-2T_1+\sum_{i=1}^N Q_i  .\eqno(46)
$$
From this equation we obtain $N$-soliton solution:

$$
g[N]=T\exp \left.\left(\int^{\xi}[U+rT_{1\xi}-2T_1+\sum_i^N Q_i]d\xi\right.
\right)g^{(N)}    .    \eqno(47)
$$

The approach which is equivalent developed above, can be realized by
the transition in (9) to the Hermitian objects. Putting
$\Phi=\Psi^{*}$, we can rewritten (9) as

$$
\Phi_{\xi}=-{\overline \Lm}\Phi U \;\;, \;\;
\Phi_{\eta}=-{\overline \Lp}\Phi V .\eqno(48)
$$

The condition of the compatibility of this system again gives equation (1)
and, consequently, we also can obtain corresponding dressing relations,
which are equivalent to ones already find. In this case for DT it should be taken
($\tilde
\Phi \equiv \Phi[-1]$):

$$
\tilde \Phi=\Phi-M_1^{*}\Phi L_1^{*} .\eqno(49)
$$

{\bf 5.} Taking into account (2), the equation (1) reduces to

$$
{\bf g}_{tt}-\frac {1}{r}(r{\bf g})_r+({\bf g}^2_t-{\bf g}^2_r){\bf g}=0  .
\eqno(50)
$$
It can be obtain by minimization an action with the Lagrangian density

$$
{\cal L}=(1/4)rTr(g_{\xi}g_{\eta})=(1/4)rTr(g_t^2-g_r^2)=
(1/2)r({\bf g}^2_t-{\bf g}^2_r) .
$$
This allow us to introduce the Hamiltonian
description of our system with Hamiltonian
($\mu$ - Lagrange multiplicator):

$$
H=\frac {1}{2}\int drr \left.\left[ ({\bf p}^2+{{\bf g}_r}^2)
-\mu ({\bf g}^2-1) \right.\right ] ,
\eqno(51)
$$
where ${\bf p}={\bf g}_t$, moreover, the phase space of the model forms by
variables ${\bf g}(r,t),\: {\bf p}(r,t)$.

Appropriate Hamiltonian equations have the form:

$$
{\bf g}_t=\frac {1}{r}\frac {\delta H}{\delta {\bf p}}=
\left.\left \{ H, {\bf g}\right.\right \}   ,
$$
$$
\eqno(52)
$$
$$
{\bf p}_t=-\frac {1}{r}\frac {\delta H}{\delta {\bf g}}=
\left.\left \{ H, {\bf p}\right.\right \}  .
$$

Here the Poissonian structure on the phase space, as in [8], is introduced
with the help of the fundamental brackets, obtained with the account
Dirac's connections
$g_ig_i-1=0,\; g_ig_{ri}=g_ig_{ti}=0$:

$$
\left.\left \{g_i(r),g_k(r')\right.\right \}=0  ,
$$
$$
\left.\left \{p_i(r), p_k(r')\right.\right \}=-\left.\left [p_i(r)
g_k(r)-p_k(r)g_i(r) \right.\right ] \delta(r-r')  ,
$$
$$
\eqno(53)
$$
$$
\left.\left \{p_i(r), g_k(r')\right.\right \}=\left.\left [\delta_{ik}-
g_i(r)g_k(r)\right.\right]\delta(r-r')   .
$$
Then for two arbitrary smooth functionals $F$ and $G$ we will have:

$$
\left.\left \{F,G \right.\right \}=
\int dr \Bigg \{\left.\left [\frac {\ddF}{\ddp}
\frac {\ddG}{\ddg}-\frac {\ddF}{\ddg}\frac {\ddG}{\ddp}\right.\right ]+
$$
$$
\eqno(54)
$$
$$
+\left.\left (
\frac {\ddG}{\ddp}{\bf g}\right.\right ) \left.\left [ \left.\left (
\frac {\ddF}{\ddg}{\bf g} \right.\right )-\left.\left (\frac {\ddF}{\ddp}
{\bf p} \right.\right ) \right.\right ]+\left.\left (\frac {\ddF}{\ddp}
{\bf g}\right.\right ) \left.\left [\left.\left (\frac {\ddG}{\ddp}
{\bf p}\right.\right)-\left.\left (\frac {\ddG}{\ddg}{\bf g}
\right.\right)
\right.\right ]\Bigg \}   .
$$

In terms the currents Hamiltonian takes the form:
$$
H=\int dr r Tr \left.\left \{U^2+V^2 \right.\right\}   . \eqno(55)
$$

It can checked by direct calculation, that the equation (1) is equivalent
to two ones:

$$
U_{\eta}=-\frac {1}{2} \left.\left [U,V \right.\right]-
\frac {1}{2r}(U-V)  ,
$$
$$
\eqno(56)
$$
$$
V_{\xi}=\frac {1}{2} \left.\left [U,V \right.\right]-
\frac {1}{2r}(U-V)   .
$$
Besides that, in terms of the variables $U,\: V$ we have one more useful
entry (1) having evident kind of the conservation law:

$$
\left.\left (U+\int^{\eta} \left.\left (\ln r \right.\right)_{\eta '} U
d {\eta} ' \right.\right )_{\eta}+\left.\left (V+\int^{\xi} \left.\left
(\ln r \right.\right )_{\xi '}V d{\xi} ' \right.\right )_{\xi} =0 . \eqno(57)
$$

{\bf 6.} Of some interest is the question about the correspondence of
model (1) and
equation of the type $\sin$-Gordon. To obtain appropriate equation let us invoke
the transformation of Pohlmayer [9]. Set

$$
{\bf g}_{\xi \xi}=c_1{\bf g}+c_2{\bf g}_{\xi}+c_3{\bf g}_{\eta}\;\;,\;\;
{\bf g}_{\eta \eta}=c_4{\bf g}+c_5{\bf g}_{\xi}+c_6{\bf g}_{\eta} ,\eqno(58)
$$
where $c_i=c_i(\xi,\eta),\; i=1, 6$ are real functions to be determined,
and introduce also the function $f=f(u)$:

$$
f(u(\xi,\eta))={\bf g}_{\xi}{\bf g}_{\eta}  .\eqno(59)
$$
Note, that, generally speaking, without loss the community, we can
suppose, that

$$
{\bf g}_{\xi}^2={\bf g}_{\eta}^2=1  .\eqno(60)
$$
Using (56)-(57), we find:

$$
{\bf g}_{\xi \xi}=-{\bf g}-\frac {ff_{\xi}+(1/2r)f(1-f)}{1-f^2} {\bf g}_{\xi}
+\frac {f_{\xi}+(1/2r)(1-f)}{1-f^2} {\bf g}_{\eta}   ,
$$
$$
\eqno(61)
$$
$$
{\bf g}_{\eta \eta}=-{\bf g}+\frac {f_{\eta}-
(1/2r)(1-f)}{1-f^2} {\bf g}_{\xi}
+\frac {-ff_{\eta}+(1/2r)f(1-f)}{1-f^2} {\bf g}_{\eta}  .
$$
Taking into account equation (1), rewritten in the vectorial form:
$$
{\bf g}_{\xi \eta}=-({\bf g}_{\xi}{\bf g}_{\eta}){\bf g}
-\frac {1}{2r}({\bf g}_{\xi}-{\bf g}_{\eta}) , \eqno(62)
$$
and using (60)-(62) we obtain the equation for the function
$u \; (f(\xi,\eta)=
\cos u)$:

$$
u_{\xi \eta}+\sin u=\frac {1}{4r} \frac {u_{\eta} -u_{\xi}}{\cos^2 \frac
{u}{2}}-\frac {1}{8r^2}\tanh \frac {u}{2} (13-\tanh^2 \frac {u}{2}) .\eqno(63)
$$
In terms of the variables $r\:,\:t$ it became:

$$
u_{tt}-u_{rr}+\sin u=\frac {1}{4r} \frac {u_r}{\cos^2 \frac
{u}{2}}-\frac {1}{8r^2}\tanh \frac {u}{2} (13-\tanh^2 \frac {u}{2})  .\eqno(64)
$$

Thus, we received the deformation of the standard equation
$\sin$-Gordon. On would suggest that the equation
(63) (or (64)) is integrable.

The alternative approach can be conclude as next.
We represent the derivatives to the currents as [10,11]:

$$
U_{\xi}=a_1 C+a_2 \left.\left [U, C \right.\right] \;\;,\;\;
V_{\eta}=b_1 C+b_2 \left.\left [V, C \right.\right] ,\eqno(65)
$$
where $C=\left.\left [U, V \right.\right]$, and coefficients $a_i,\: b_i$ are
functions $\xi$ and $\eta$. Multiplying the first equation of the system
(65) on $V$, and second equation on
$U$ and calculating the traces, we have ($\tau
\equiv -(1/2) Tr(UV) = \cos u$):

$$
a_2=-\frac {{\tau}_{\xi}}{4(1-\tau^2)}-\frac {1}{8r}\frac {1}{1+\tau},
$$
$$
\eqno(66)
$$
$$
b_2=\frac {\tau_{\eta}}{4(1-\tau^2)}+\frac {1}{8\tau}\frac {1}{1+\tau}  .
$$
Putting $U=i U_i \sigma_i, \; V=i V_i \sigma_i$, turn out from
(61) to the vectorial system and multiplying the first equation in terms of
scalar on  ${\bf V}_{\xi}$, and second on ${\bf U}_{\eta}$,
we find coefficients $a_1$ and $b_1$:
$$
a_1=\frac {{\bf U}_{\xi}{\bf V}_{\xi}}{2{\sin^2 u}} +\frac {a_2}{r} , \; \;
\; b_1=-\frac {{\bf U}_{\eta}{\bf V}_{\eta}}{2{\sin^2 u}}-\frac {b_2}{r} .
\eqno(67)
$$

{\bf 7.} In conclusion we note, that approach, proposed in this paper,
can be easy transferred on the case vacuum equations of Ernst in
the theory of gravitation [12]. Then the role of matrices $g$ will
play the symmetrical and real matrices of nondiagonal part of the
metrics, the involutions and kinematical
connections (14)-(20) will be changed also.

The author is grateful to M.A.Salle for the attentions to this work and the useful
discussions.

This work was supported by RFBR (grant N 00-01-00-480).

\vskip2.5cm
\centerline {\bf {References}}
\vskip1cm
[1]. \parbox[t]{12.7cm}
{{ Salle M.A and Matveev V.B. Darboux Transformation and Solitons (1991),
Springer-Verlag.}}
\vskip0.3cm
[2]. \parbox[t]{12.7cm}
{{ Mikhailov A.V. and Yaremchuk A.I. Nucl.Phys.{\bf B202}, 508(1982).}}

\vskip0.3cm
[3]. \parbox[t]{12.7cm}
{{  Belinskii V.A. and Zakharov V.E. Journal Experimentalnoi i Teoreticheskoi
phiziki {\bf 77}, 3(1979)(in Russian).}}

\vskip0.3cm
[4]. \parbox[t]{12.7cm}
{{  Lipovski V.D. and Shirokov A.V. Zapiski nauchnych seminarov LOMI {\bf 209},
 150(1994)}(in Russian).}

\vskip0.3cm
[5]. \parbox[t]{12.7cm}
{{ Gutshabash E.Sh. and Lipovskii V.D. Zapiski nauchnych seminarov LOMI {\bf 199},
  71(1992)}(in Russian).}
\vskip0.3cm
[6]. \parbox[t]{12.7cm}
{{  Gutshabash E.Sh., Lipovskii V.D. and Nikulichev C.C. Teoreticheskaja i
Matematicheskaja phizika {\bf 115}, 323(1998)};  solv-int 9900142.}

\vskip0.3cm
[7]. \parbox[t]{12.7cm}
{{  Matos T. Matematicheskie zametki {\bf 58}, 710(1995)(in Russian).}}

\vskip0.3cm
[8]. \parbox[t]{12.7cm}
{{ Taktadjan L.A. and Faddeev L.D. The Hamiltonian Approach in the Theory of
Solitons, 1986, Moscow, Nauka (in Russian).}}
\vskip0.3cm
[9]. \parbox[t]{12.7cm}
{{  Pohlmayer K. Commun.Math.Phys. {\bf 46}, 207(1976).}}

\vskip0.3cm
[10]. \parbox[t]{12.7cm}
{{ Zakharov V.E. and Mikhailov A.V. Journal Experimentalnoi i Teoreticheskoi
Phiziki {\bf 74}, 1953(1978)(in Russian).}}

\vskip0.3cm
[11]. \parbox[t]{12.7cm}
{{ Vekslerchik V.E. J.Phys.A {\bf 27}, 6299(1994).}}

\vskip0.3cm
[12]. \parbox[t]{12.7cm}
{{ Korotkin D.A. and Matveev V.B. Algebra i Analiz {\bf 1}, 77 (1989)}(in Russian).}
\end{document}